\documentclass[runningheads]{llncs}
\usepackage{color}
\usepackage{graphicx}
\usepackage{amssymb}
\usepackage[plainpages=false, pdfborderstyle={/S/U/W 0}]{hyperref}
\usepackage{fancyvrb}
\usepackage{ifpdf}
\usepackage{listings}
\usepackage{xypic}
\usepackage{bussproofs}

\renewcommand{\verb}{\lstinline}

\begin{document}

\title{Smart matching}

\author{Andrea Asperti, Enrico Tassi}
\institute{
 Department of Computer Science, University of Bologna \\
 \email{asperti@cs.unibo.it}
 \and
 Microsoft Research-INRIA Joint Center\\
 \email{enrico.tassi@inria.fr}
}

\date{}
\maketitle

\begin{abstract}
One of the most annoying aspects in the formalization of
mathematics is the need of transforming notions to match a given, 
existing result. 
This kind of transformations, often based on a conspicuous
background knowledge in the given scientific domain (mostly
expressed in the form of equalities or isomorphisms), are 
usually implicit in the mathematical discourse, and it would 
be highly desirable to obtain a similar behaviour in interactive provers. 
The paper describes the superposition-based implementation
of this feature inside the Matita interactive theorem prover, 
focusing in particular on the so called {\em smart application} tactic, 
supporting smart matching between a goal and a given result.
\end{abstract}

\section{Introduction}
\let\thefootnote\relax\footnotetext{The final publication of this paper is available at www.springerlink.com}
The mathematical language has a deep contextual nature, whose 
interpretation often presupposes not trivial skills 
in the given mathematical discipline. The most common and 
typical example of these ``logical abuses'' is the implicit 
use of equalities and isomorphisms, allowing a mathematician to 
freely move between different incarnations of a same entity in a 
completely implicit way. Equipping ITP systems with the
capability of reasoning up to equality yields an essential 
improvement of their intelligence, making the communication 
between the user and the machine sensibly easier. 

Techniques for equational reasoning have been broadly investigated
n the realm of automated theorem proving 
(see eg \cite{BG94,paramodulation,equality-handbook}). 
The main deductive mechanism is a {\em completion}
technique \cite{Knuth-Bendix} attempting to transform a given set of 
equations into a confluent rewriting system so that two terms 
are equal if and only if they have identical normal forms. 
Not every equational theory can be presented as a confluent
rewriting system, but one can progressively approximate it 
by means of a refutationally complete method called 
{\em ordered completion}.
The deductive inference rule used in completion procedures is 
called {\em superposition}:
it consists of first unifying one side of one equation with a subterm 
of another, and hence rewriting it with the other side.
The selection of the two terms to be unified is guided by a suitable 
{\em term ordering}, constraining inferences and sensibly pruning the 
search space. 

Although we are not aware of any work explicitly focused on 
superposition techniques for interactive provers, 
the integration between fully 
automatic provers (usually covering paramodulation)
and interactive ones is a major 
research challenge and many efforts have been already done in this
direction: for instance, KIV has been integrated with the tableau 
prover $3T^AP$ \cite{AB98}; HOL has been integrated with various 
first order provers, such as Gandalf \cite{Hurd99} and Metis; 
Coq has been integrated with Bliksem \cite{BHN02}; 
Isabelle was first integrated with a purpose-built prover 
\cite{blast} and more recently with Vampire \cite{MQP06}. 
The problems of these integrations are usually of two kinds:
(a) there is a {\em technical} difficulty in the forward and backward 
translation of the information between systems, due to the different
underlying logics (ITP systems are usually higher-order, and some of
them intuitionistic); 
(b) there is a {\em pragmatical} problem in the management 
of the knowledge base
to be used by the automatic solver, since it can be huge
(so we cannot pass it at every invocation), and it
grows dynamically (hence, it cannot be exported in advance).

A good point of the superposition calculus (and not the
last reason for restricting the attention to this important
fragment) is that point (a), in this context, becomes relatively trivial
(and the translation particularly effective).
As for point (b), its main consequence is that the communication
between the Interactive Prover and the Problem Solver, in order
to be efficient, cannot be {\em stateless}: the two systems must 
share a common knowledge base. This fact, joined with the
freedom to adapt the superposition tool to any possible 
specific requirement of the Matita system convinced us to 
rewrite our own solver, instead of trying to interface Matita 
with some available tool.
This paper discusses our experience of implementation of a 
(first order) superposition calculus (Section \ref{sec:superposition}), 
its integration within the (higher-order) Matita interactive prover
\cite{matita-jar-uitp} (Section \ref{sec:integration}),
and in particular its use for the implementation of a
{\em smart application} tactic, 
supporting smart matching between a goal and a given results
(Section \ref{sec:applyS}). We shall conclude with a large number of
examples of concrete use of this tactic.  

\section{The Matita superposition tool}\label{sec:superposition}
One of the components of the automation support provided by
the Matita interactive theorem prover is a first order, 
untyped superposition tool. 
This is a quite small and compact application (little
more than 3000 lines of OCaml code), well separated by the rest 
of the system. It was entirely rewritten during the summer 2009
starting from a previous prototype (some of whose functionalities
had been outlined in \cite{hopr}), 
with the aim to improve both its abstraction and performance. 
The tool took part to the 22nd CADE ATP System Competition, 
in the unit equality division, scoring in fourth position,
beating glorious systems such as Otter or Metis \cite{Metis}, 
and being awarded as the best new entrant tool of the competion
\cite{Sutcliffe09}.

In the rest of this section we shall give an outline, as concise
as possible, of the theory and the architecture of the tool. 
This is important in order to understand its integration with
the interactive prover.

\subsection{The superposition calculus in a nutshell}
Let $\mathcal{F}$ bet a countable alphabet of functional symbols, and 
$\mathcal{V}$ a countable alphabet of variables. 
We denote with $\mathcal{T}(\mathcal{F},\mathcal{V})$  the set of terms 
over $\mathcal{F}$ with variables in $\mathcal{V}$. A term 
$t\in \mathcal{T}(\mathcal{F},\mathcal{V})$ is either a 0-arity element of 
$\mathcal{F}$ (constant), an element of $\mathcal{V}$ (variable), 
or an expression of the form $f(t_1, \dots, t_n)$ where $f$ is a element of 
$\mathcal{F}$ of arity $n$ and $t_1, \dots, t_n$ are terms.

Let $s$ and $r$ be two terms. 
$s|_p$ denotes the subterm of $s$ at position $p$ and $s[r]_p$
denotes the term $s$ where the subterm at position $p$ has been 
replaced by $r$.

A substitution is a mapping from variables to terms. 
Two terms $s$ and $t$ are unifiable if there exists a substitution
$\sigma$ such that $s\sigma = t\sigma$. In the previous case,
$\sigma$ is called a most general unifier (mgu) of $s$ and $t$ if 
for all substitution $\theta$ such that
$s\theta = t\theta$, there exists a substitution $\tau$ which satisfies
$\theta = \tau \circ \sigma$. 

A literal is either an abstract predicate (represented by a term), 
or an equality between two terms. A clause $\Gamma \vdash \Delta$ 
is a pair of multisets of literals: the negative literals $\Gamma$,
and the positive
ones $\Delta$. If $\Gamma = \emptyset$ (resp. $\Delta = \emptyset$), 
the clause is said to be positive (resp. negative).

A Horn clause is a clause with at most one positive literal. 
A unit clause is a clause
composed of a single literal. A unit equality is a unit clause 
where the literal is an equality. 

A strict ordering $\prec$ over $\mathcal{T}(\mathcal{F},\mathcal{V})$
is a transitive and irreflexive (possibly
partial) binary relation. An ordering is {\em stable} under substitution
if $s \prec t$ implies $s\sigma \prec t\sigma$ for all terms $t, s$ and
substitutions $\sigma$. A well founded monotonic ordering stable under
substitution is called {\em reduction ordering} (see \cite{Dershowitz82}).
The intuition behind the use of reduction orderings for 
limiting the combinatorial explosion of new equations during
inference, is to only rewrite big terms to smaller ones.

\begin{figure}[hbt]
\[
\begin{array}{ccc}
\quad\mbox{{\bf superposition left}}\quad & 
\quad\mbox{{\bf superposition right}}\quad &
\quad\mbox{{\bf equality resolution}}\quad \\\\
  \begin{array}{c}
  \vdash l = r \quad\quad t_1 = t_2 \vdash\\
  \hline
  (t_1[r]_p = t_2 \vdash)\sigma
  \end{array}
&
  \begin{array}{c}
      \vdash l = r \quad\quad \vdash t_1 = t_2 \\\hline
      (t_1[r]_p = t_2 \vdash)\sigma
  \end{array}
&
\begin{array}{c}
    t_1 = t_2 \vdash\\\hline
      \Box
  \end{array} \\
 \multicolumn{2}{c}{\mbox{if }
\sigma = mgu(l, {t_1}|_p), t_1|_p \neq x, l\sigma\not\preceq r\sigma \mbox{ and } t_1\sigma \not\preceq t_2\sigma} 
& \mbox{if } \exists\sigma = mgu(t_1, t_2).
\end{array}
\]
\caption{Inference rules}
\end{figure}


For efficiency reasons, the calculus must be integrated with a few
additional optimization rules, the most important one being
demodulation (\cite{demodulation}).
\begin{figure}[hbt]
\[
\begin{array}{ccc}
\quad\mbox{{\bf subsumption}}\quad & 
\quad\mbox{{\bf tautology elimination}}\quad &
\quad\mbox{{\bf demodulation}}\quad \\\\
  \begin{array}{c}
    S \cup \{C, D\}\\\hline
    S \cup \{C\}

  \end{array}
&
  \begin{array}{c}
    S \cup \{\vdash t = t\} \\\hline
    S
  \end{array}
&
\begin{array}{c}
    S \cup \{\vdash l = r, C\} \\\hline
    S \cup \{\vdash l = r, C[r\sigma]_p\}
  \end{array} \\
\mbox{if }\exists\sigma, D\sigma \equiv C &
&
\mbox{if }l\sigma \equiv C|_p \mbox{ and } l\sigma \succ r\sigma
\end{array}
\]
\caption{Simplification rules}
\end{figure}

\subsection{The main algorithm}
\label{sec:main_algorithm}
A naive implementation of the superposition calculus could just
combine (superpose) all known clauses in all (admitted) ways, and repeat
that process until the desired clause (called \emph{goal}) is resolved.
To avoid useless duplication of work, it is convenient to keep clauses
in two distinct sets, traditionally called {\em active} and {\em
passive}, with the general invariant that clauses in the active set
have been already composed together in all possible ways.  At every
step, some clauses are selected from the passive set and added to the
active set, then superposed with the active set, and consequently with
themselves (\emph{inference}). 
Finally, the newly generated clauses are added to the
passive set (possibly after a simplification). 

A natural selection strategy, resulting in a very predictable
behaviour, would consist in selecting the whole passive set at each
iteration, in the spirit of a breadth first search.  Unfortunately the
number of new equations generated at each step grows extremely fast,
in practice preventing the iteratation of the main loop more than a few
times.
  
To avoid this problem, all modern theorem provers 
(see e.g. \cite{vampire_annals}) adopt the opposite solution.
According to some heuristics, like size and goal similarity for example, 
they select only {\em one} passive clause at each step. Not to loose
completeness, some fairness conditions are taken into account (i.e.
every passive clause will be eventually selected). 
\begin{figure}[htp]
\begin{center}
\includegraphics[width=0.55\textwidth]{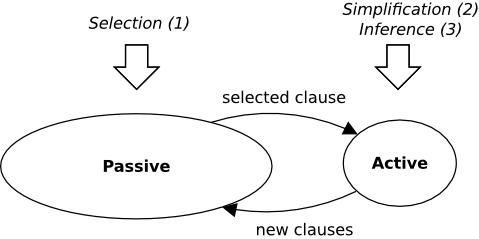}
\caption{given-clause loop\label{mainloop}}
\small{Numbers in parentheses reflect the steps order.}
\end{center}
\end{figure}
This approach falls under the name \emph{given-clause} (Figure
\ref{mainloop}), and its main advantage is that the passive set grows much
slower, allowing a more focused and deeper inspection of the search
space that consequently allows to find proofs that require a much higher
number of main loop iterations.

The main drawback of this approach is that it makes the procedure way more
sensible to the selection heuristics, leading to an essentially
unpredictable behaviour.

\subsection{Performance issues}
In order to obtain a state-of-the-art tool able to compete with
the best available systems one has eventually to take into account
a lot of optimizations and techniques developed for this purpose
during the last thirty years. 

In the following we shall shortly describe the most critical areas,
and, for each of them, the approach adopted in Matita.




\subsubsection{Orderings used to orientate rewriting rules}
On complex problems (e.g. problems in the TPTP library with rating
greater then $0.30$) the choice of a good ordering for inference rules
is of critical importance. 
We have implemented several orderings,
comprising standard Knuth-Bendix (KBO), non recursive Knuth-Bendix
(NRKBO), lexicographic path ordering (LPO) and recursive path ordering
(RPO).  The best suited ordering heavily depends on the kind of
problem, and is hard to predict: our approach for the CADE ATP
System Competition was to run in parallel different processes with
different orderings. 

On simpler problems (of the kind required for the smart application
tactic of section \ref{sec:examples}), the given-clause algorithm is 
less sensitive to the term-ordering, and we may indifferently 
choose our preferred strategy, opportunely tuning the library 
(we are currently relying on LPO). 

\subsubsection{Selection strategy}

The selection strategy currently implemented by Matita is a based on
combination of age and weight. The weight is a positive integer that 
provides an estimation of the ``complexity'' of the
clause, and is tightly related to the number of occurrences of symbols 
in it. 

Since we are not interested in generating (counter) models of
false statements, we renounced to be complete, and we silently drop
inferred clauses that would slow down the main loop too much due to
their excessive size. 

Another similar optimization we did not implement but we could
consider as a future development is Limited Resource Strategy
\cite{LRS}, which basically allows the procedure to skip some
inference steps if the resulting clauses are unlikely to be processed,
mainly because of a lack of time.

\subsubsection{Data structures and code optimization}
We adopted relatively simple data structures (like discrimination
\cite{McCune} trees for term indexing), and a purely functional (in
the sense of functional programming) implementation of them. After
some code optimisation, we reached a point where very fast functions
are the most expensive, because of the number of calls (implied by the
number of clauses), even if they operate on simple data structures.

Since we are quite satisfied with the actual performance, we did not
invest resources in adopting better data structures, but we believe
that further optimizations will probably require implementing more
elaborate data structures, such as substitution \cite{Graf95} or
context trees \cite{codedcontexttrees}, or even adopt an indexing
technique that works modulo associativity and
commutativity~\cite{assoccommut}, that looks very promising 
when working on algebraic structures.

\subsubsection{Demodulation}

Another important issue for performance is demodulation: the given clause
algorithm spends most of its time (up to 80\%) in simplification, hence
any improvement in this part of the code has a deep impact on performance.
However, while reduction strategies, sharing issues and abstract machines
have been extensively investigated for lambda calculus (and in general
for left linear systems) less is known for general first 
order rewriting systems. 
In particular, while an innermost (eager) 
reduction strategy seem to work generally better than an outermost 
one (especially when combined with lexicographic path ordering), one 
could easily create examples showing an opposite behaviour (even 
supposing to always reduce needed redexes). 

\section{Integrating superposition with Matita}
\label{sec:integration}

\subsection{Library management}
A possible approach to the integration of superposition with Matita
is to solve all goals assuming that all equations part of the library lie in 
the passive set, augmented on the fly with the equations in the local
context of the ongoing proof. 

The big drawback of this approach is that, starting essentially from
the same set of passive equations at each invocation on a different
goal (differing only for the local context), the given clause algorithm would 
mostly repeat the same selection and composition operations 
over and over again. 
It is clear that, if we wish to superpose library equations, this operation
should not be done at run time but in background, 
once and for all. Then we have to face a dual problem, namely to
understand when stopping the saturation of the library with new 
equations, preventing an annoying pollution with trivial results
that could have very nasty effects for selection and memory 
occupation. We would eventually like to have mechanisms to drive
the saturation process.

A natural compromise is to look at library equations not as a passive set, 
but as the \emph{active} one. 
This means that every time a new (unit) equation is added to the 
library it also goes through one main given-clause loop, as if it was 
the newly selected passive equation: it is simplified, composed
with all existing active equations (i.e. all other equations in the library,
up to simplification), and the newly created equations
are added to the passive list. At run time, we shall then strongly
privilege selection of local equations or goals.

This way, we have a natural, 
simple but traceable syntax to drive the saturation process, by 
just listing in library the selected equations.
As a side effect, this approach reduces the verbosity of the library by
making it unnecessary to declare (and name explicitly) trivial variants
of available results that are automatically generated by superposition.

\subsection{Interfacing CIC and the superposition engine}
Our superposition tool is first order and untyped, while the
Matita interactive prover is based on a variant of the Calculus
of Inductive Construction (CIC), a complex higher-order intuitionistic
logical systems with dependent types. The communication between
the two components is hence far from trivial.   

Instead of attempting a complex, faithful encoding of CIC in 
first order logic (that is essentially the approach adopted
for HOL in \cite{MP08}) we choose to follow a more naif
approach, based on a forgetful translation that remove types
and just keeps the first order applicative skeleton of CIC-terms.

In the opposite direction, we try to reconstruct the missing 
information by just exploiting the sophisticated inference
capability of the Matita {\em refiner}~\cite{hints}, that is 
the tool in charge of transforming the user input
into a machine understandable low-level CIC term.

Automation is thus a best effort service, in the sense
that  not only it may obviously fail to produce a proof, but sometimes
it could produce an argument that Matita will fail to understand, 
independently from the fact if the delivered proof was ``correct''
or less.

The choice to deal with untyped first order equations in the superposition
tool was mostly done for simplicity and modularity reasons.  
Moving towards a typed setting would require a much tighter integration 
between the superposition tool and the whole system, due to
the complexity of typing and unification, but does not seem to pose
any major theoretical problem.

\subsubsection{The forgetful encoding}
Equations $r =_T s$ of the calculus of constructions
are translated to first order equations by merely following the
applicative structure of $r$ and $s$, and translating 
any other subterm into an opaque constant. The type $T$ of the equation
is recorded, but we are not supposed to be able to compute types
for subterms.

In spite of the fact of neglecting types, the risk of 
producing ``ill-typed'' terms via superposition rules is moderate.
Consider for instance the superposition left rule (the reasoning is similar
for the other rules)
  \begin{displaymath}
    \frac{
      \vdash l = r \quad\quad t_1 = t_2 \vdash
    }{
      (t_1[r]_p = t_2 \vdash)\sigma
    }
  \end{displaymath}
where $\sigma = mgu(l, {t_1}|_p)$ and $l\sigma \not\preceq r\sigma$.
The risk is that $t_1|_p$ has 
a different type from $l$, resulting into an illegal rewriting
step. Note however that $l$ and $r$ are usually rigid terms, whose
type is uniquely determined by the outermost symbol. Moreover, 
$t_1|_p$ cannot be a variable, hence they must share this outermost
symbol. If $l$ is not rigid, it is usually a variable $x$ and if 
$x \in r$ (like e.g. in $x=x+0$) we have (in most orderings)
$l \preceq r$ that again rules out rewriting in the wrong direction.

This leads us to the following notion of {\em admissibility}. 
We say that an applicative term $f(x_1,\dots,x_n)$ is {\em implicitly
typed} if its type is uniquely determined by the type of $f$. 
We say that an equation $l = r$ is admissible if both $l$ and $r$
are implicitly typed, or $l \preceq r$ and $r$ is implicitly typed.
Non admissible equations are not taken into account by the superposition 
tool\footnote{A more liberal, but also slightly more expensive solution 
consists in indexing any equation and systematically try to read back each 
result of a superposition step in CIC, dropping it if it is not 
understood by the refiner.}.

In practice, most unit equalities are admissible.
A typical counter example is an equation of the kind
$\forall x,y:unit. x = y$, where $unit$ is a singleton type.

On the other side, non-unit equalities are often not admissible.
For instance, a clause of the kind $x \wedge y = true \vdash x = true$
could be used to rewrite any term to true, generating meaningless,
ill typed clauses. 
Extending superposition beyond the unit equality
case does eventually require to take types into consideration.


\subsection{(Re)construction of the proof term}\label{sec:proofs}
Translating a first-order resolution proof into a higher-order
logic natural deduction proof is a notoriously difficult issue,
even more delicate in case of intuitionistic systems, as the one
supported by Matita. While resolution {\em per se}
is a perfectly constructive process, skolemization and
transformation into conjunctive normal forms are based on
classical principles. 

Our choice of focusing on the 
superposition calculus was also motivated by the fact it poses less 
difficulties, since skolemization is not needed and thus proofs have a 
rather simple intuitionistic interpretation.

Our technique for reconstructing a proof term relies as much as possible
on the refinement capabilities of Matita, in particular for
inferring implicit types.
In the superposition module, each proof step is encoded as a 
tuple
\begin{verbatim}
  Step of rule * int * int * direction * position * substitution
\end{verbatim}
where \verb+rule+ is the kind of rule which has been applied,
the two integers are the two $id's$ of the composing equations
(referring to a ``bag'' of unit clauses),
\verb+direction+ is the direction the second equation is applied
to the first one, \verb+position+ is a path inside the rewritten term and
finally \verb+substitution+ is the mgu required for the rewriting step.

Every superposition step is encoded by one of the following terms:
\[
\begin{array}{rl}
eq\_ind\_l: & \forall A: \textsc{Type}. \forall x:A. \forall P: A \to
  \textsc{Prop}.P~x \to \forall y: A. x = y \to P~y\\
eq\_ind\_r: & \forall A: \textsc{Type}. \forall x:A. \forall P: A \to
  \textsc{Prop}.P~x \to \forall y: A. y = x \to P~y
\end{array}
\]
where left (\verb+_l+) and right (\verb+_r+) must be understood w.r.t.
backward application, and where \verb+P+ is the one hole context that 
represents the position in which the superposition occurred.

    At the end of the superposition procedure, if a proof is found, 
either a trivial goal has been generated, or a fact subsumes one of 
the active goals. In that latter case, we perform a rewriting step on 
the subsumed goal, so that we fall back into the
previous case.
    Thus, when the procedure successfully stops, the selected clause 
is of the form $s = t$ where $s$ and $t$ are unifiable. 
We call it the meeting point, because forward steps
(superposition right) and backward steps (superposition left) 
meet together when this trivial
clause is generated, to compose the resulting proof.
To generate a CIC proof term, the clauses are topologically sorted, 
their free variables are explicitly quantified, and nested let-in 
patterns are used to build the proof. 

The most delicate point of the translation is closing each clause
w.r.t. its free variables, since we should infer a type for them, and since
CIC is an explicitly polymorphic language it is often the case that
the order of abstractions does matter (e.g. variables standing for types
must in general be abstracted before polymorphic variables).  

The simplest solution is to generate so called ``implicit'' arguments
leaving to the Matita {\em refiner} the burden of guessing them.

For instance, superposing 
$\mathrm{lencat}: len~A~x + len~A~y = len~A~(x \underline{@} y)$ with 
$\mathrm{catA}: x @ (y @ z) \stackrel{\leftarrow}{=} (x @ y) @ z$ 
at the underlined position and in the given direction
gives rise to the following piece of code, where question marks stand
for implicit arguments:

\begin{lstlisting}
let clause_59:
   $\forall w: ?.\forall x: ?.\forall y: ?.\forall z: ?.$
      $len~w~(x @ y) + len~w~z = len~w~(x @ (y @ z))$
 $:=$ 
   $\lambda w:?.\lambda z:?.\lambda x:?.\lambda y:?.$
      eq_ind_r ($List~w$) $((x@y)@z))$
        $(\lambda hole:List~w. len~w~(x @ y) + len~w~z = len~w~hole)$
        (lencat $w~(x@y)~z$) ($x@(y@z)$) (catA $w~x~y~z$) in
\end{lstlisting}
Note that $w$ \emph{must} be abstracted first, since it occurs in the
(to be inferred) types for $x,y$ and $z$. Also note the one hole context
expressed as an anonymous function whose abstracted variable is named $hole$,
corresponding to the position of $x\underline{@}y$ in the statement of 
lencat.

%
The interesting point is that refining is a complex operation, using
e.g. hints, and possibly calling back the automation itself: the 
interpretation of the proof becomes hence a dialog between the
system and its automation components, aimed to figure out a correct
interpretation out of a rough initial trace.

A more sophisticated translation,
aimed to produce a really nice, human-readable output in the form 
of a chain of equations, 
is described in \cite{hopr}.

\section{Smart application}
\label{sec:applyS}
The most interesting application of superposition (apart from its
use for solving equational goals) is the implementation of a more
flexible application tactic. As a matter of fact, one of the most 
annoying aspects of formal development is the need of transforming 
notions to match 
a given, existing result. As explained in the introduction, most of these 
transformations are completely transparent to the typical mathematical
discourse, and we would like to obtain a similar behaviour in interactive
provers.

Given a goal $\mathcal{B}$ and a theorem t: $A \to B$, the goal is to try
to match $B$ with $\mathcal{B}$ {\em up to the available equational 
knowledge base}, in
order to apply $t$. We call it, the {\em smart application} of $t$ to 
$G$. 
\vspace{-0.2cm}
\begin{figure}[htp]
\begin{center}
\includegraphics[width=0.25\textwidth]{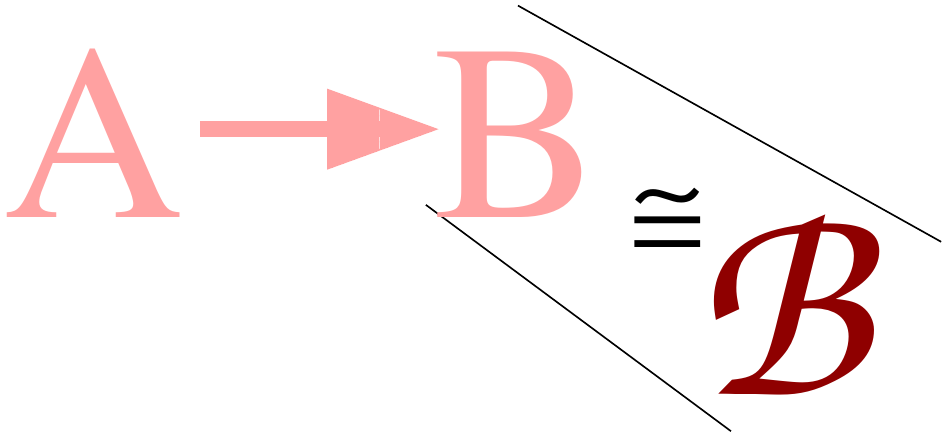}
\caption{Smart application\label{smartapp}}
\end{center}
\end{figure}
\vspace{-0.8cm}
We use superposition in the most direct way, exploiting on
one side the higher-order features of CIC, and on the other
the fact that the translation to first order terms does
not make any difference between predicates and functions:
we simply generate a
goal $B = \mathcal{B}$ and pass it to the superposition tool (actually, 
it was precisely this kind of operation that motivated our original
interest in superposition). If a proof is found, $\mathcal{B}$ is transformed
into $B$ by rewriting and $t$ is then 
normally applied.

Superposition, addressing a typically undecidable problem, 
can easily diverge, while we would like to have a reasonably 
fast answer to the smart application invocation, as for any other 
tactic of the system. We could simply
add a timeout, but we prefer to take a different, more predictable 
approach. As we already said, the overall idea is that superposition
right steps - realising the {\em saturation} of the equational
theory - should be thought of as background operations. Hence, at run
time, we should conceptually work as if we had a {\em 
confluent} rewriting system, and the only operation worth to do
is {\em narrowing} (that is, left superposition steps). Narrowing
too can be undecidable, hence we fix a given number of narrowing 
operations to apply to each goal (where the new goal instances generated at
each step are treated in parallel). The number of narrowing steps 
can be fixed by the user, but a really small number is usually
enough to solve the problem if a solution exists. 
\vspace{-0.3cm}
\section{Examples}
\label{sec:examples}

\begin{example}
\label{example:smart1}
Suppose we wish to prove that the successor function
is le-reflecting, namely
\[(*)\hspace{.5cm}\forall n,m. S n \le S m \to n \le m\]
Suppose we already proved that the predecessor function is monotonic:
\[monotonic\_pred: \forall n,m. n \le m \to pred\; n \le pred\; m\]
We would like to merely ``apply'' the latter to prove the former.
Just relying on unification, this would not be possible, 
since there is no way to match
$pred\; X \le pred\; Y$ versus $n \le m$ unless {\em narrowing} the 
former. By superposing twice with the equation 
$\forall n. pred (S n) = n$ we 
immediately solve our matching problem via the substitution
$\{X := S n, Y := S m\}$.  Hence, the smart application of
$monotonic\_pred$ to the goal $n \le m$ succeeds, opening the new
goal $S n \le S m$ that is the assumption in $(*)$.
\end{example}

\begin{example}
\label{example:smart2}
Suppose we wish to prove $n \le m*n$ for all natural numbers 
$n,m$. Suppose we already proved that multiplication 
is left-monotonic, namely
\[monotonic\_le\_times\_l: \forall n,a,b. a \le b \to a*n \le b*n \]
In order to apply this result, the system has to find a suitable
$?_a$ such that $?_a*n = n$, that is easily provided by the
identity law for times.
\end{example}

\begin{example}
\label{example:smart3}
In many cases, we just have local equational variants of the
needed results. Suppose for instance we proved that multiplication
in le-reflecting in its right parameter:
\[le\_times\_to\_le\_times\_r: \forall a,n,m. a * n \le a * m \to n \le m \]
Since times is commutative, this also trivially implies the left version:
\[monotonic\_le\_times\_l: \forall a,n,m. n * a \le m *a \to n \le m \]
Formally, suppose to have the goal  $n \le m$ under the assumption
$(H)\; n*a \le m*a$. By applying $le\_times\_to\_le\_times\_r$ 
we obtain a new goal $?_a * n \le ?_a * m$ that is a smart variant of
$H$. 
\end{example}

\begin{example}
\label{example:smart4}
Suppose we wish to prove that 
$(H)\; a*(S n) \le a*(S m)$ implies $a * n \le a*m$, where $S$ 
is the successor function (this is a subcase in the inductive
proof that the product by a positive constant $a$ is le-reflecting). 
Suppose we already proved that
the sum is le-reflecting in its second argument:
\[le\_plus\_to\_le\_plus\_r: \forall a,n,m. a + n \le a + m \to n \le m \]
By applying this result we obtain the new goal
$?_a + a*n \le ?_a + a*m$, and if we have the expected equations 
for times, we can close the proof by a smart application of $H$.
\end{example}

\begin{example}
\label{example:smart5}
Consider the goal $n < 2*m$ under the assumptions $(H)\; 0 < m$ and
$(H1)\; n \le m$. Suppose that we defined $x < y$ as $x + 1 \le y$.
Morevoer, by the defining equation of times we should 
know something like $2*m = m + (m + 0)$.\footnote{The precise
shape depends by the specific equations available on times.}
Hence the goal is equal to $n+1 \le m + (m + 0)$, and the idea 
is to use again the
monotonicity of plus (in both arguments):
\[le\_plus\;n\;m: \forall a,b. n \le m \to a \le b \to n+a \le m+b \]
The smart application of this term to the goal $n < 2*m$ succeeds,
generating the two subgoals $n \le m$ and $1 \le m + 0$. The former
one is the assumption $H1$, while the latter is a {\em smart} variant
of $H$. 
\end{example}


\begin{example}
\label{example:smart7}
Let us make an example inspired by the theory of programming
languages. Suppose to have a typing relation $\Gamma \vdash M:N$
stating that in the environment $\Gamma$ the term $M$ has type $N$.
If we work in De Bruijn notation, the weakening rule requires 
lifting\footnote{The lifting operation $\uparrow^n\!\!(M)$ is meant to
relocate the term $M$ under $n$ additional levels of bindings: 
in other words, it
increases by $n$ all free variables in $M$.}
%
 
\[\mbox{weak}: \Gamma \vdash M:N \to 
\Gamma,A \vdash \uparrow^1\!\!(M) : \; \uparrow^1\!\!(N)\]
Suppose now we have an axiom stating that $ \vdash * : \Box$ where 
$*$ and $\Box$ are two given sorts. We would like to generalize
the previous result to an arbitrary (legal) context $\Gamma$.
To prove this, we have just to apply weakenings (reasoning by induction 
on $\Gamma$). However, the normal application of $weak$ would fail, 
since the system should be able to guess two terms $M$ and $N$ such 
$\uparrow^1\!\!(M) = *$ and  $\uparrow^1\!\!(N) = \Box$. 
If we know that for any constant $c$, $\uparrow^1\!\!(c) = c$ (that comes from
the definition of lifting) we may use such an equation to enable the
smart application of $weak$.
\end{example}


\subsubsection{Performance}
In Figure~\ref{times} we give the execution times for the
examples of smart applications discussed in the previous section
(in bytecode). Considering these times, it is important to stress again 
that the smart application tactics
does not take any hint about the equations it is supposed 
to use to solve the matching problem, but exploits all the equations
available in the (imported sections of the) library.

The important point is that smart application is fast enough to
not disturb the interactive dialog with the proof assistant, 
while providing a much higher degree of flexibility than the
traditional application.

\begin{figure}
\[
\begin{array}{|c|l|l|}
\hline
\mbox{example} & \mbox{applied term} & \mbox{execution time} \\\hline 
1              &  momonotonic\_pred  & 0.16 s. \\\hline
2              &  momonotonic\_le\_times\_l  & 0.23 s.\\\hline
3              &  H:a*n \le a*m    & 0.22 s.\\\hline
4              &  H:a*(S n) \le a*(S m)    & 0.15 s.\\\hline
5              &  le\_plus\; n\; m & 0.57 s.\\\hline
6              &  weak & 0.15 s. \\\hline
\end{array}
\]
\caption{\label{times}Smart application execution times}
\vspace{-0.5cm}
\end{figure}
\section{Related works and systems}
Matita was essentially conceived as a light version of Coq \cite{Coq},
sharing the same foundational logic (the Calculus of
Inductive Constructions) and being partially compatible with it
(see \cite{ck-sadhana} for a discussion of the main differences between 
the two systems at kernel level). Hence, Coq is also the most natural
touchstone for our work. The {\tt auto} tactic of Coq does not perform
rewriting; this is only done by a couple of specialized tactics, called
{\tt auto rewrite} and {\tt congruence}. The first tactic 
carries out rewritings according to sets of oriented equational rules
explicitly passed as arguments to the tactic (and previously
build by the user with suitable vernacular commands). Each rewriting 
rule in some base is applied to the goal until no further reduction
is possible. The tactic does not perform narrowing, nor any form of
completion. 
The {\tt congruence} tactic implements the standard Nelson and Oppen 
congruence closure algorithm \cite{congruence}, which is a decision 
procedure for {\em ground} equalities with uninterpreted symbols; 
the Coq tactic only deals with equalities in the local context.
Both Coq tactics are sensibly weaker than superposition that 
seems to provide a good surrogate for several decision procedures
for various theories, as well as a simple framework for composing 
them (see e.g \cite{armandoCSL}). 

Comparing the integration of superposition in Matita with similar
functionalities provided by Isabelle is twofold complex, due
not only to the different approaches, but also to the different
underlying logics. 

In Isabelle, equational reasoning can be both delegated to external tools 
or dealt with internally by the so called {\em simplifier}. 
Some of the the external tools Isabelle is interfaced with provide
full support to paramodulation (and hence superposition), but
the integration with them is stateless, possibly requiring to pass 
hundreads of theorems (all the current visible environment) at each
invocation. In Matita, the active set is \emph{persistent}, and grows 
as the user proves new equations.\\
Of more interest is the comparison with Isabelle's internal
simplifier. The integration of this tool with the library is manual: only
lemmas explicitly labelled and oriented by the user are taken into account 
by the simplifier. Moreover, these lemmas are only used to demodulate and 
are not combined together to infer new rewriting rules. 
Nevertheless, a pre-processing phase allows the user to label theorems 
whose shape is not an equation. For example a conjunction of 
two equations is interpreted as two distinct rewriting rules, or a negative
statement $\lnot A$ is understood as $A = False$. 
The simplifier is also able to take into account guarded
equations as long as their premises can be solved by the simplifier itself.
Finally it detects equations that cannot be oriented by the user, like
commutativity, and restricts their application according to the demodulation
rule using a predefined lexicographic order.\\
Anyway, the main difference from the user's perspective comes from a 
deep reason that has little to do with the simplifier or any other
implemented machinery. Since Isabelle is based on classical logic,
co-implication can be expressed as an equality. Hence, in Isabelle
we can prove much more equations at the prositional level and use
them for rewriting. Any concrete comparison between the two provers 
with respect to equational reasoning is thus
inherently biased, since many problems encountered in one system would look
meaningless, artificial or trivial when transposed into the other one.

\section{Conclusions}
\label{sec:conclusions}
We described in this paper the ``smart'' application tactic
of the Matita interactive theorem prover. The tactics
allow the backward application of a theorem to a goal, 
where matching is done up to the data base of all equations 
available in the library. The implementation of the tactics
relies on a compact superposition tool, whose architecture
and integration within Matita have been described in the first 
sections. The tool is already performant (it was awarded best new 
entrant tool at the 22nd CADE ATP System Competition)
but many improvements can still be done for efficiency, 
such as the implementation of more sophisticated data structures 
for indexes (we currently use discrimination trees).

Another interesting research direction is to extend the management
of equality to setoid rewriting \cite{setoidrewriting}. 
Indeed, the current version of the superposition tool just works
with an intensional equality, and it would be
interesting to try to figure out how to handle more general 
binary relations. The hard problem is proof reconstruction, but
again it seems possible to exploit the sophisticated capabilities
of the Matita refiner \cite{hints} to automatically check the legality of
the rewriting operation (i.e. the monotonicity of the context
inside which rewriting has to be performed), exploiting some of the
ideas outlined in \cite{damaJFR}. 

One of the most promising uses of smart application is inside
the backward-based automation tactic of Matita. In fact, smart
application allows a smooth integration of equational reasoning
with the prolog-like backward applicative mechanisms that, according to 
our first experimentations looks extremely promising. As a 
matter of fact, the weakest point of smart application is that 
it does not relieve the user form the effort of finding the ``right''
theorems in the library or of guessing/remembering their names 
(although it allows to sensibly reduce the need of variants of 
a given statement in the repository). 
A suitably constrained automation tactic could 
entirely replace the user in the quest of candidates for the smart
application tactic. Since searching is a relatively expensive operation, 
the idea is to ask the automation
tactic to return an explicit trace of the resulting proof
(essentially, a sequence of smart applications) to
speed-up its re-execution during script development.

\smallskip
\noindent
{\bf Acknowledgements} We would like to thank Alberto Griggio and
Maxime D\'en\`es for their contribution to the 
implementation of the superposition tool.

\end{document}